\documentstyle[amssymb,prl,aps,multicol]{revtex}
\begin{document}
\title{Kondo effect in optics of quantum dots}
\author{Konstantin Kikoin and Yshai Avishai}
\address{Department of Physics, Ben-Gurion University, Beer-Sheva 84 105,
Israel}
\date{\today}
\maketitle

\begin{abstract}
Anderson impurity model for semiconductor quantum dot is extended to
take into account both particle and hole branches of charge excitations. It
is shown that in dots with even number of electrons where the Kondo effect
is
absent in the ground state, novel midgap excitonic states emerge in the
energy spectrum due to Kondo-type shake-up processes. The relevance of the
model to rare earth ions adsorbed on metallic surfaces is discussed.
\end{abstract}


\begin{multicols}{2}

The Kondo effect was formulated initially for a spin 1/2 impurity coupled to
metallic environment. Later on, various realizations of this effect were
studied in more complicated physical systems. Examples are the Kondo effect
influenced by crystal field excitations, the multichannel Kondo problem,
electron scattering by two-level systems, etc \cite{Zawad98}. Usually, the
scattering center is represented by its spin (pseudo-spin) degrees of
freedom. In this work we consider yet another generalization of the Anderson
impurity model, where the impurity possesses both electron and hole branches
of single-particle excitations as well as two-particle (excitonic) states.
Discrete excitonic states do not show up in a conventional situation of an
impurity immersed in a conduction band, even when such an impurity has
internal charge degrees of freedom (e.g., rare-earth impurities with two
unfilled electron shells \cite{Kibur93}). However, long-living excitons can
exist when the scattering center is spatially isolated  from the conduction
electrons, like in semiconductor quantum dots coupled by tunneling to
metallic
leads \cite{Tsu91} or in atoms adsorbed on metallic surfaces \cite{Grim}. It
is well known \cite{Glazr88a} that the Kondo scattering could results in
resonant tunnel current. Recently the Kondo features were observed
in the conductance of quantum dots \cite{Gogo98}.
We demonstrate in this work that Kondo-type shake-up process
leads to an exciton formation in a quantum dot with even number of electrons
(where the Kondo effect is absent in the ground state).

To reveal new qualitative features which appear in the Kondo problem due to
excitons, we consider the following model of semiconductor quantum dot:
\begin{equation}
H=H_{d}+H_{b}+H_{t}=H_{0}+H_{t}.
\label{1.1}
\end{equation}
The first term, $H_{d}$, describes the isolated dot
\begin{equation}
H_{d}=E_{c}n_c-E_{v}n_v+
\frac{U_c}{2}n_c(n_c-1) + \frac{U_v}{2} n_v(n_v-1)
,  \label{1.2}
\end{equation}
where $\sigma $ is the
spin projection, $n_{c}=\sum_{\sigma }d^\dagger_{c\sigma }d_{c\sigma }$ is
the
occupation number for the electrons at an empty conduction ($c$) level,
$n_{v}=\sum_{\sigma }\overline{n}_{v\sigma } =
\sum_{\sigma}d_{v\sigma }d^\dagger_{v\sigma }$
is the occupation number for holes at
a filled valence level $E_v< \varepsilon_F$. We define the gap as an
excitation energy of the electron-hole pair,  
$\Delta=E_{c}-E_{v}$. The band electrons in the leads
with the Fermi energy  $\varepsilon_F$ and the tunneling between
the leads and the dot with the tunneling amplitudes $V_{k(c,v)}$
 are described by the second and third terms in
(\ref{1.1}) respectively,
\begin{equation}
H_{b}=\sum_{k\sigma }\varepsilon _{k}c_{k\sigma }^{\dagger }c_{k\sigma
},\;\;H_{t}=\sum_{k\sigma }\sum_{j=c,v}(V_{kj}c_{k\sigma }^{\dagger
}d_{j\sigma }+\text{H.c.})
\label{1.3}
\end{equation}

No Kondo coupling is expected in the ground state $\Psi_G$, and the
strong Hubbard repulsion $U_{c,v}$
suppresses double occupation of electron or hole levels, so that only
the states with singly charged dot
\begin{equation}
|kc\rangle =\frac{1}{\sqrt{2}}\sum_{\sigma }d_{c\sigma }^{\dagger}
c_{k\sigma }|0\rangle ,\;\;
|kv\rangle =\frac{1}{\sqrt{2}}\sum_{\sigma}c_{k\sigma }^{\dagger }
d_{v\sigma }|0\rangle ,
\label{1.6}
\end{equation}
are admixed by
tunneling to the neutral
state $|0\rangle$ (the filled Fermi sea and valence level $E_v$) of
the isolated dot, and one can consider this admixture as a small
first order perturbation (spin 1/2 is considered).
There is also no Kondo effect in photoexcitation
spectra 
$$D^0 +h\nu \to D^\pm +e,h$$ 
where $D^0, D^\pm$ are the
ground state 
of the dot, and the charged states with a hole $v$ and excess electron
$c$,
respectively;   
$e (h)$ stands for the electron (hole)
promoted to the leads due to absorption of the light quantum $h\nu$.
We will show, however, that the Kondo-like processes develop in the
spectrum of electron-hole excitations of the quantum dot given by the
operator
\begin{equation}
\left| e\right\rangle =\frac{1}{\sqrt{2}}\sum_{\sigma }d_{c\sigma }^{\dagger
}d_{v\sigma }|0\rangle \equiv B^{\dagger }\left| 0\right\rangle .
\label{1.2a}
\end{equation}
These states can be excited by means of photon absorption/emission processes
which are described by the Hamiltonian
$H'=\sum_{ij,\sigma}P_{ij}d_{i\sigma }^{\dagger
}d_{j\sigma }\exp(-i\omega t)$, where
$P_{ij}$ is the matrix element of the dipole operator $\hat{P}$.

The optical line shape is given by the Kubo-Greenwood formula ,
\begin{equation}
W(h\nu )\simeq
{\rm Im} \frac{1}{\pi }\langle i|\hat{P}\hat{R}(h\nu )\hat{P}| i\rangle
\label{1.2b}
\end{equation}
where $\hat{R}(z)=(z - H)^{-1}$, and $\langle i|\ldots| i\rangle$ means
averaging of initial state over the equilibrium ensemble.
Due to Hubbard exclusion mechanism only the states $|0\rangle$
survive in the brackets of response function
(\ref{1.2b}), and one has to calculate the retarded excitonic
Green function
\begin{equation}
G^{R}_{ee}(z) =-i\int dte^{izt}\theta(t)\langle
| [B^\dagger (t)B(0)]|\rangle
\label{2.5a}
\end{equation}
Employing the operator
identity $\hat{R}=\hat{R}_{0}+\hat{R}_{0}H_{t}\hat{R}$
for the resolvent $\hat{R}$
we construct the
perturbation series in $H_t$ where the first term is
$R_{ee}=\langle e|\hat{R}|e\rangle$.
Having in mind the fact that the strong coupling limit
$T\ll T_K$ hardly can be achieved (see below),
we restrict ourselves by the weak coupling limit of $T$ exceeding the
Kondo temperature $T_K$.

In a simplest approximation one neglects the intermediate states with
multiple electron-hole pairs in the leads
(see, e.g., \cite{Lacr81}).
Then the closed system of equations for the matrix
elements  $R_{se}^{-}=\langle s|\hat{R}^{-}|e\rangle $ where
$\langle s|$ are the states admixed to the exciton by the tunneling
interaction, is obtained.
The structure of the
system is illustrated in fig.1, and its solution for the excitonic Green
function has the form
$R_{ee}^{-}(\epsilon )=[z -\Sigma _{0}(\epsilon)]
[{\cal D}(\epsilon )]^{-1}.$
The poles of this function are determined by the equation
\begin{equation}
{\cal D}(z)=\det \left|
\begin{tabular}{cc}
$z -\Sigma _{0}(z)$ & $\quad -\Sigma _{0e}(z)$ \\
$-\Sigma _{0e}(z)\quad $ & $z -\Delta -\Sigma _{e}(z)$
\end{tabular}
\right| =0.
\label{2.10}
\end{equation}
The self energies are given by
\begin{eqnarray}
\Sigma_{0} & = & \Sigma_{cc}^{+}+\Sigma_{vv}^{-},\;\;
2\Sigma_{e}  =  \Sigma_{vv}^{+}+\Sigma_{cc}{-},\nonumber \\
\sqrt{2}\Sigma _{0e} & = &
\Sigma_{cv}^{-} -\Sigma_{vc}^{+}
\label{1.13}
\end{eqnarray}
where
\begin{equation}
\Sigma_{jl}^{+}=
\sum_{k}\frac{w^*_jw_lf_k}{z -E_{c}+\varepsilon _{k}},\;
\Sigma_{jl}^{-}=
\sum_{k}\frac{w^*_jw_l\bar{f}_k}{z +E_{v}-\varepsilon _{k}},
\label{sigmaP}
\end{equation}
$f_k$ is the equilibrium distribution function for lead electrons,
$\bar{f}_k=1-f_k$.
 The tunneling matrix elements $w$ are defined as
$\langle 0|H_{t}|kc\rangle \approx \sqrt{2}\langle e|H_{t}|kv\rangle \approx
w_{c}~,\;\;\langle 0|H_{t}|kv\rangle \approx -\sqrt{2}\langle
e|H_{t}|kc\rangle \approx w_{v}^{\ast }.$ 
Then integration gives 
\begin{equation}
{\rm Re}\Sigma^\pm_{jl}\approx
\frac{\Gamma_{jl}}{2\pi }\ln
\frac{(\epsilon -\Delta_{c,v})^2+(\pi T)^2}{D^2},
\label{sigmaM}
\end{equation}
where $\Delta_c=E_{c}-\epsilon _{F}$, $\Delta_v=\epsilon _{F}-E_v$,
$\Gamma_{jl}=\pi \rho w^{*}_jw_l$,
$\rho $ is the density of states at the Fermi
surface, $\varepsilon _{F}$ is the Fermi energy and $D$ is the conduction
electron bandwidth. The signes $\pm$ correspond to c,v, respectively,
$z=\epsilon+is$.

As was mentioned above, the renormalization of the {\it ground state}
at $\epsilon\sim 0$ is a trivial second-order effect in the case
of $E_{c}>\epsilon _{F}$, $E_{v}<\epsilon _{F}$ (even number of
electrons in the ground state of the dot). However,
novel features appear in the {\it excitation spectrum} at
$\epsilon\sim \Delta_{c}, \Delta_{v}$. In a process of tunnel relaxation
the electron and the hole with finite energy $\epsilon$
induce Kondo-like peaks at the Fermi level.
The physics of "Kondo-excitons" can be illustrated by considering two
limiting cases.\noindent\\
(i) Symmetric configuration $w_c=w_v\equiv w$,
$\Delta_c=\Delta_v \equiv \Delta/2$.
In this case $\Sigma _{0e}$ vanishes identically,
and the secular equation becomes
\begin{equation}
\epsilon -\Delta -\Sigma _{e}(\epsilon )=0\; .
\label{2.17}
\end{equation}
It has a solution at
$\epsilon \sim \widetilde{\Delta}=\Delta+\Sigma_e(\Delta) $, corresponding
to the normal exciton. Besides,
${\rm Re}\Sigma _{e}(\epsilon )$ diverges at $T\to T_K$,
\begin{equation}
T_{K}=D\exp \left( -\pi\Delta/ 2\Gamma \right) \; .
\label{2.14}
\end{equation}
As a result a peak arises in ${\rm Im} G_{ee}$ at $\epsilon$ around
$\Delta/2$
which corresponds to maximum of light absorption/emission at this energies.
When the particle-hole symmetry is slightly violated ($\delta \neq 0$),
$\Delta_c =\Delta /2-\delta $, $\Delta_v=\Delta /2+\delta $,
$\delta \ll \Delta /2$, this midgap peak disappears with encreasing $\delta$
due to cancellation
of singular terms with opposite sign in eq. (\ref{2.10}). We will see
below that this state is also fragile against the lifetime effect.
\noindent\\
(ii) Strongly asymmetric configurations,
$\Delta_c\ll \Delta_v$, $w_c\gg w_v$. This case is closer to
the situation in real systems where the electrons
are usually less localized then the holes. In the extreme case the level
$E_v$ is below the bottom of the band $\varepsilon_k$, 
("Schrieffer-Wolff (SW) limit" \cite{Gun83}), whereas the electron
is in resonance with the band
electrons. The secular equation (\ref{2.10}) may be rewritten as
\begin{equation}
2(\epsilon -\Delta) =
\Sigma_{vv} ^{+}+\Sigma_{cc} ^{-}+ \frac
{|\Sigma_{cv} ^{-}-\Sigma_{vc} ^{+}|^2}
{(\epsilon-\Sigma_{cc} ^{+}-\Sigma_{vv}^{-})}
\label{sec}
\end{equation}
To find the resonance solution at $\epsilon\sim\Delta_v$, we neglect
the smooth contributions $\Sigma_{ij}^+(\epsilon)$ in comparison with the
singular self energies  $\Sigma_{ij}^-(\epsilon)$, and use approximate
value of $\Sigma_{vv}^{-}(\Delta_v)\approx -\eta\Delta_c $
in the denominator of the ratio in r.h.s. $(\eta=(w_v/w_c)^2\ll 1)$.
Then we have 
\begin{equation}
2(\epsilon -\Delta) \approx
\Sigma_{cc} ^{-}+
{\eta |\Sigma_{cc} ^{-}|^2}(
{\Delta_v+\eta \Delta_c})^{-1}\;.
\label{seca}
\end{equation}
Like eq. (\ref{2.17}), this equation has a Kondo-like pole
at $\epsilon=\Delta_v, T=\widetilde{T}_K$, where
$\widetilde{T}_K=D\exp (-2\pi\Delta_c/\widetilde{\Gamma}_{cc})$ and
$\widetilde{\Gamma}_{cc} \approx \Gamma_{cc}(1-\eta\Delta_c/\Delta_v)$.
Repeating the procedure for the resonance at $\epsilon\sim \Delta_c$,
we leave in (\ref{sec}) only the terms $\Sigma_{ij}^+$  and find the
midgap peak at $\epsilon=\Delta_c$,
with $\widetilde{T}_K=D\exp (-2\pi\Delta_v/\widetilde{\Gamma}_{vv})$,
$\widetilde{\Gamma}_{vv}\approx \Gamma_{vv}\eta\Delta_c/\Delta_v$.

Thus, the Kondo-type processes
manifest themselves as a shake-up effect with a shake-up energy of 
$\Delta_{v,c}$, i.e.
as a final state
interaction between the $(e,h)$ pair in the dot and the Fermi continuum
in the lead.
The $T$-dependent logarithmic singularity in excitonic self energy is a
precursor of
"orthogonality catastrophe" in close analogy with the corresponding
anomaly in a $d$-electron self energy in the conventional Anderson model 
\cite{Lacr81}.  
In the latter case the
Kondo peak transforms to undamped Abrikosov-Suhl resonance in a ground
state \cite{Gun83}. However, this is not the case for the
Kondo exciton because of the
finite lifetime $\tau_l$ of the $(e,h)$ pair.
The most important contributions to $\tau_l$ are given by the same tunneling
processes which are responsible for the very existence of the midgap states.
To take them into account one should include the states with
$(e,h)$ pairs in the leads in the Green function expansion. These
states
appear in 4th order of the perturbation theory. In a non-crossing
approximation (NCA) they result in renormalization of the self energies
\cite{Gun83,Bickers},
\begin{eqnarray}
\widetilde{\Sigma}_{jl} ^{+}(\epsilon) & = & \frac{\Gamma_{jl}}{\pi }
\int_{-D}^D
\frac{f(\varepsilon)d\varepsilon}{\epsilon-E_{c}+\varepsilon-
B^{-}_{jl}(\epsilon)}, \nonumber \\
\widetilde{\Sigma}_{jl}^{-}(\epsilon) & = & \frac{\Gamma_{jl}}{\pi }
\int^{D}_{-D}
\frac{\bar{f}(\varepsilon)d\varepsilon}{\epsilon+E_{v}-\varepsilon-
B_{jl}^{-}(\epsilon)}.
\label{1.15}
\end{eqnarray}
where $B_{jl}^{\pm}$ are the integrals similar to
(\ref{sigmaP}). In particular,
\begin{eqnarray}
B_{cc}^{+} & = &
\int_{-D}^D\frac{f(\varepsilon)d\varepsilon'}{\pi}\left[
\frac{\Gamma_{vv}}{\epsilon -\varepsilon' +\varepsilon}+
\frac{2\Gamma_{cc}}{\epsilon -\varepsilon'
+\varepsilon-\Delta}\right]  \nonumber\\
B_{vv}^{-} & = &
\int^D_{-D}\frac{\bar{f}(\varepsilon)d\varepsilon'}{\pi}\left[
\frac{\Gamma_{cc}}{\epsilon+\varepsilon'-\varepsilon}+
\frac{2\Gamma_{vv}}{\epsilon+\varepsilon'-\varepsilon+\Delta}\right]
\label{1.16}
\end{eqnarray}

In a symmetric case (\ref{2.17}) the self energy
$B^{\pm}(\epsilon)$ has imaginary part $\approx 2\Gamma$
at $\epsilon \approx \Delta/2$. Thus, the Kondo processes initiated by
one of the partners in the electron-hole pair are killed by the damping
of its counterpart due to the same tunneling mechanism.
More interesting is the asymmetric configuration described by eq.
(\ref{sec}).
In this case the singularity of
$\widetilde{\Sigma}_{cc}^-$ in electron channel at
$\epsilon\sim \Delta_v$ survives in a SW limit for a hole
[$E_v$ below the bottom of conduction band,
${\rm Im}B^{+}_{cc}(\varepsilon_F)=0$],
whereas the electron lifetime
given by ${\rm Im} B_{vv}^{-}\sim \Gamma_{cc}$ kills the hole peak
like in the symmetric case. The electron midgap state survives also when
$E_v$ has a finite width, provided $\Gamma_{vv}<\widetilde{T}_K$, i.e.
$2\pi\Delta_c/\Gamma_{cc}<\ln D/\Gamma_{vv}.$

Thus, in the case (ii) the main
peak of optical transition at $h\nu =\widetilde{\Delta }$ is accompanied by
a satellite peak at $h\nu \approx \Delta_v $. The form 
of this peak is determined by eq. (\ref{1.2b}), i.e., by 
\begin{equation}
\frac{1}{\pi}{\rm Im} G_{ee}(h\nu)=
\frac{1}{\pi}{\rm Im} \left[
h\nu-\Delta-\widetilde{\Sigma}^-_{ee}(h\nu)+i\hbar\tau_l^{-1}\right]
^{-1}.
\label{2.19}
\end{equation}
where $\widetilde{\Sigma}^-_{ee}$ is given by the r.h.s. of eq.
(\ref{seca}).The lineshape $W(h\nu)$ strongly depends on $T/\tilde{T}_K$ 
and $\tau_l$ \cite{Kibur93,Lacr81}.

Two systems where this theory can be applied are suggested below.
The first one is an
ensemble of semiconductor quantum dots (e.g., the 
nanosize Si clusters embedded in
amorphous SiO$_{2}$ matrix \cite{Tsu91}). Luminescence of confined excitons
was observed in these clusters \cite{Min96}, and the tunneling current
through this system exhibits the Coulomb blockade effect \cite{Tsu91,Nic93}.
In the Coulomb blockade regime the energy $U$ which enters $H_{d}$, is
introduced as $U=E_{i}^{(n+1)}-E_{i}^{(n)}+e^{2}/C_{eff},$ where
$E_{i}^{(n)} $ is the energy level of the $n$-th quantum state of the
electron (hole) in the empty well ($n=1$ in our case), and $C_{eff}$ is
an effective capacitance of the barrier layers.
Experimental \cite{Nic93} and
theoretical \cite{Bab92} estimates of $C_{eff}$ for Si
nanoclusters give $U\approx 0.1\div 0.3$ eV, hence the general condition for
realization of the Kondo effect, $\Gamma \ll |E_{i}|,E_{i}+U,$ is 
satisfied in these systems. However, the experimental value of $U$ is small
compared with
$\Delta $ $\sim 1.2\div 1.3$ eV \cite{Bab97}. Taking into
account finite $U$ means inclusion of doubly occupied states
$\left|2c\right\rangle $, $\left| 2v\right\rangle $
in the set (\ref{1.6}), (\ref{1.2a}).
In close analogy with the conventional Anderson model \cite{Gun83} one
expects redistribution of the spectral weight of neutral states $|0\rangle $
and $|e\rangle $ in favor of the states $|2c\rangle $, $|2v\rangle $, and
increase of $T_{K}$ with decreasing $U$. However, the inequality
$T_{K}\ll \Delta $ ensures the existence
of the midgap states.

Another important condition for the appearance of midgap states is
long enough excitonic lifetime $\tau _{e}\gg \hbar /T_{K}$.
In addition to tunneling contribution $\Gamma$ discussed above, one should
also take into account the electron-hole recombination in the dot resulting
in a width $\gamma$ of the excitonic level. This damping gives
the contribution to imaginary part $\sim \Gamma\gamma/\pi\Delta\ll\Gamma$
in eqs. (\ref{2.17}) and (\ref{sec}) for a Kondo pole.
In any case, the
experimentally estimated value is $\tau _{e}\sim 10^{-6}$ s
for the singlet exciton \cite
{Brong98}. Therefore the Kondo-type processes can survive in these systems
if $T_{K}\gg 10^{-9}$ eV, which is a realistic condition.
We therefore believe that our model qualitatively describes
those properties of Si nanoclusters which play a crucial role
in the formation of Kondo-type states in  optical spectra.

The second possible realization is mixed valent rare earth atoms
adsorbed on a metallic surface. It is known that the Anderson
model can be applied to adatoms with strongly interacting
electrons (see\cite{Grim} for a review). In this case $H_{t}$
(\ref{1.3}) corresponds to covalent bonding between an adsorbed
atom and a substrate, $V$ is the corresponding hybridization
integral between the electrons in adatom and those in the nearest
sites of the metallic surface layer, $U$ is the intra-atomic
Coulomb repulsion which prevents charging of the adatom in the
process of chemisorption. The model was originally proposed for
hydrogen atoms adsorbed on surfaces of transition metals
\cite{New69}. Later on, the possibility of Kondo-type spin
polarization of substrate electrons around adatom spin in the case
of $U\gg V^{2}/D$ was discussed \cite{Schrief74}. The most
promising candidates from the point of view of excitonic effects
are the adatoms with unstable valence, e.g., Sm, whose ground
state electronic configuration is $4f^{6}6s^{2}$. The Sm atoms can
be adsorbed on surfaces of transition metals (Ni, Co, Cu, Mo). In
the process of adsorption, Sm loses its $s$-electrons and exists
in two charged states Sm$^{2+}$ and Sm$^{3+}$ depending on the
concentration of Sm ions on the surface. In particular, the
isolated ions Sm$^{2+}(4f^{6})$ are observed on Mo surface at low
submonolayer coverage \cite{Sten89}. The unfilled 4f shell forms a
resonant f-state close to the Fermi level of the metal. The
excited 5d state forms another level above $\varepsilon _{F}$.
Thus we arrive at a two-level system described by our Hamiltonian
$H_{d}$ (\ref{1.2}). Since the ground state term of the
configuration $4f^{6}$ is a singlet $^{7}F_{0}$, one cannot expect
the Kondo coupling for such adatom. However, in the course of
virtual transitions between the adatom and the substrate the
states $|kv\rangle $ and $|kc\rangle $ arise with excess electron $e_k$
above
$\varepsilon_F$ (configuration $4f^{5}e_k$) and
a hole below $\varepsilon_F$ (configuration
 $4f^{5}5dh_k$). According to our calculations, one can excite not
only the conventional atomic excited state with energy $\Delta
=E(4f^{5}5d)-E(4f^{6})$ but also the midgap states with energy
close to $\Delta ^{\prime }=E(4f^{5}e_{F})-E(4f^{6})$ where
$e_{F}$ stands for the electron on the Fermi level of the
substrate.

To summarize, this work suggests a generalization of the Anderson impurity
model which takes into account excitonic degrees of freedom. The model
exhibits Kondo effect in the excited state, despite its absence in the
ground state. It is predicted that satellite excitonic peaks of a Kondo
origin can be seen in the optical spectra of semiconductor quantum dots or
rare earth atoms adsorbed on metal surfaces.

We thank A. Polman and M. Brongersma for stimulating discussions of the
optical properties of Si nano-clusters. This research is supported by
Israeli Academy grants "Nonlinear Current Response of Multilevel Quantum
Systems" and "Strongly Correlated Electron Systems in Restricted
Geometries".

\end{multicols}

{\bf {\Large Figure Caption}}

\noindent {\bf {\large Fig. 1}}. Building blocks of the Green  function
expansion and the secular equation
(\ref{2.10}). The arrows indicate the tunneling processes
which connect different states of this set.


\begin{references}
\bibitem{Zawad98}  D.L. Cox and A. Zawadowski, Adv. Phys., {\bf 47}, 599
(1998).

\bibitem{Kibur93}  K.A. Kikoin and S.B. Burkatovskii, {\it \ }Journ.  
Moscow Phys. Soc.{\bf 3}, 139 (1993).

\bibitem{Tsu91}  Q. Ye et al, Phys. Rev. B{\bf 44}, 1806 (1991).

\bibitem{Grim}  T.L. Einstein et al, in {\it Theory of Chemisorption},
edited by J.R. Smith (Springer, Berlin, 1980), p.184.

\bibitem{Glazr88a}  L.I. Glazman and M.E. Raikh, JETP Lett. {\bf 47}, 452,
(1988); T.K. Ng and P.A. Lee, Phys. Rev. Lett., {\bf 61}, 1768 (1988).

\bibitem{Gogo98}  D. Goldhaber-Gordon et al, Nature {\bf 391}, 156 (1998);
S.M. Cronenwett et al, Science, {\bf 281}, 540 (1998).

\bibitem{Lacr81}  C. Lacroix, Journ. Phys. F{\bf 11}, 2389 (1981);
N.S. Wingreen and Y. Meir, Phys. Rev. B{\bf 49}, 11040 (1994).

\bibitem{Gun83}  A.C. Hewson, {\it The Kondo Problem to Heavy Fermions},
Cambridge, University Press, 1993.

\bibitem{Min96}  K.S. Min et al, Appl. Phys. Lett. {\bf 69}, 2033 (1996).

\bibitem{Bickers} P. Coleman, Phys. Rev. B{\bf29}, 3035 (1984);
N. Bickers, Rev. Mod. Phys. {\bf 59}, 845 (1987).

\bibitem{Nic93}  E.N. Nicollian and R. Tsu, J. Appl. Phys. {\bf 74}, 4020
(1993).

\bibitem{Bab92}  D. Babic et al, Phys. Rev. B{\bf 45}, 14150 (1992).

\bibitem{Bab97}  D. Babic and R. Tsu, Superlattices and Microstruct.
{\bf 22}, 581 (1997).

\bibitem{Brong98}  M. Brongersma, Ph. D. Thesis, Utrecht, 1998.

\bibitem{New69}  D.M. Newns, Phys. Rev. {\bf 178}, 1123 (1969).

\bibitem{Schrief74}  J.R. Schrieffer, in {\it Dynamic Aspects of Surface
Physics}, Rend. Scuola Enrico Fermi, Varenna, v. 58 (1974).

\bibitem{Sten89}  A. Stenborg et al, Phys. Rev. B{\bf 40}, 5916 (1989).
\end{references}
\end{document}